# Linear Complementarity Algorithms for Infinite Games

John Fearnley[1], Marcin Jurdziński[1], and Rahul Savani[2]

[1] Department of Computer Science, University of Warwick, UK
[2] Department of Computer Science, University of Liverpool, UK

**Abstract.** The performance of two pivoting algorithms, due to Lemke and Cottle and Dantzig, is studied on linear complementarity problems (LCPs) that arise from infinite games, such as parity, average-reward, and discounted games. The algorithms have not been previously studied in the context of infinite games, and they offer alternatives to the classical strategy-improvement algorithms. The two algorithms are described purely in terms of discounted games, thus bypassing the reduction from the games to LCPs, and hence facilitating a better understanding of the algorithms when applied to games. A family of parity games is given, on which both algorithms run in exponential time, indicating that in the worst case they perform no better for parity, average-reward, or discounted games than they do for general P-matrix LCPs.

## 1 Introduction

In this paper we consider infinite-duration zero-sum games played on finite graphs, such as parity, average-reward, and discounted games. Parity games are important in the theory of algorithmic formal verification because they provide a combinatorial characterization of the meaning of nested inductive and coinductive definitions, as formalized in the modal $\mu$-calculus and other fixpoint logics [12]. In particular, deciding the winner in parity games is polynomial-time equivalent to checking non-emptiness of non-deterministic parity tree automata, and to the modal $\mu$-calculus model checking, two fundamental algorithmic problems in automata theory, logic, and verification [7, 18, 12]. Discounted and average-reward games have been introduced by Shapley [17] and Gillette [11] in the 1950s, and they have been extensively studied in the game theory, mathematical programming, algorithms, and AI communities [21, 8]. Parity, average-reward, and discounted games have an intriguing complexity-theoretic status. The problems of deciding the winner in these games are some of the few known combinatorial problems in NP $\cap$ co-NP (and even UP $\cap$ co-UP [13]) that are not known to be solvable in polynomial time.

The linear complementarity problem (LCP) is a fundamental problem in mathematical programming. It naturally captures equilibrium problems, as well as the complementary slackness and Karush-Kuhn-Tucker conditions of linear and quadratic programming, respectively. The monograph of Cottle *et. al.* [6] is the authoritative source on the LCP. In general, deciding if an LCP has a solution is NP-complete [3]. If, however, the matrix (which is a part of the LCP input) is a P-matrix (i.e., if all its principal minors are positive) then the problem is arguably easier computationally. Every P-matrix LCP (P-LCP) has a unique solution and computing it is in PLS ∩ PPAD. A significant amount of effort has been invested by the mathematical programming community towards finding an efficient algorithm for the P-LCP, which has led to a wide body of literature in this area. Polynomial-time reductions from simple stochastic games [10, 19] and discounted games [14] to the P-LCP have been recently proposed, however the techniques commonly used to solve P-LCPs remain largely unknown in the infinite games community. It is possible that these techniques could shed new light on the computational complexity of solving infinite games.

In this paper we consider two classical pivoting algorithms for the P-LCP, Lemke's algorithm and the Cottle-Dantzig principal pivoting algorithm, and we study their performance on P-LCPs obtained from discounted games by the reduction of Jurdziński and Savani [14]. Our first main contribution is to describe both algorithms purely as a process that works on the original discounted game, bypassing the reduction from games to the P-LCP, and hence we facilitate their analysis without the need to consider or understand concepts of the LCP theory. We present the algorithms for discounted games because they have technical advantages that make the exposition particularly transparent [14]. We argue, however, that this is done without loss of generality: the algorithms can be readily applied to parity games and average-reward games because there are transparent polynomial-time reductions from parity games to average-reward games [16, 13], and from average-reward games to discounted games [21].

It has long been known that the two algorithms can take exponential time when applied to P-LCPs. However, it is not known whether these lower bounds hold for the LCPs that arise from infinite games. Our second main contribution is to prove that there is a family of discounted games on which the algorithms of Lemke, and Cottle and Dantzig run in exponential time, and hence we indicate limitations of the classical LCP theory in the context of infinite games. Our family of examples are derived from those given by Björklund and Vorobyov [2] for their strategy improvement algorithm for average-reward games. For technical convenience and without loss of generality, we present these families of hard examples as discounted games; it is easy to construct parity and average-reward games from which those discounted games are obtained via the standard reductions [16, 13, 21].

We stress that these lower bounds are not fatal. The lower bound for Lemke's algorithm requires a specific covering vector and the lower bound for the Cottle-Dantzig algorithm relies on a specific choice of ordering on the vertices. The covering vector and the ordering are free choices left up to the user of the algorithm. This situation can be compared to the classical strategy improvement algorithms for infinite games [5, 20, 2]. It has long been known that these algorithms can be made to run in exponential time by choosing sufficiently bad vertices to switch [15]. However, it has only recently been shown that reasonable switching policies can be made to run in exponential time [9]. The complexity of our algorithms when equipped with reasonable covering vectors and reasonable orderings remains open. The literature on P-LCPs contains exciting complexity results for special cases. For example Adler and Megiddo [1] studied the performance of Lemke's algorithm for the LCPs arising from linear programming problems. They showed that for randomly chosen linear programs and a carefully selected covering vector, the expected number of pivots performed by the algorithm is quadratic. Our results open the door to extending such analyses to infinite games.

## 2 Preliminaries

A binary discounted game is given by a tuple $G = (V, V_{\text{Max}}, V_{\text{Min}}, \lambda, \rho, r^\lambda, r^\rho, \beta)$, where $V$ is a set of vertices and $V_{\text{Max}}$ and $V_{\text{Min}}$ partition $V$ into the set of vertices of player Max and the set of vertices of player Min, respectively. Each vertex has exactly two outgoing edges which are given by the left and right successor functions $\lambda, \rho : V \to V$. Each edge has a reward associated with it given by the functions $r^\lambda, r^\rho : V \to \mathbb{R}$. Finally, the discount factor $\beta$ is such that $0 \leq \beta < 1$.

The game begins with a token on a starting vertex $v_0$. In each round, the player who owns the vertex on which the token is placed chooses one of the two successors of that vertex and moves the token to that successor. In this fashion the two players form an infinite path $\pi = \langle v_0, v_1, v_2, \dots \rangle$ where $v_{i+1}$ is equal to either $\lambda(v_i)$ or $\rho(v_i)$. The path yields the infinite sequence of rewards $\langle r_0, r_1, r_2, \dots \rangle$, where $r_i = r^\lambda(v_i)$ if $\lambda(v_i) = v_{i+1}$, and $r_i = r^\rho(v_i)$ otherwise. The payoff of an infinite path is denoted by $\mathcal{D}(\pi) = \sum_{i=0}^{\infty} \beta^i r^i$. Since the game is zero-sum, player Max wins $\mathcal{D}(\pi)$ and player Min loses an equal amount.

A positional strategy for player Max is a function that, for each vertex belonging to player Max, chooses one of the two successors of the vertex. The strategy is denoted by $\chi : V_{\text{Max}} \to V$ with the condition that, for every vertex $v$ in $V_{\text{Max}}$, the function $\chi(v)$ is equal to either $\lambda(v)$ or $\rho(v)$. Positional strategies for player Min are defined analogously. The sets of pure positional strategies for Max and Min are denoted by $\Pi_{\text{Max}}$ and $\Pi_{\text{Min}}$, respectively. Given a pair of positional strategies, $\chi$ and $\mu$ for Max and Min respectively, and an initial vertex $v_0$, there is a unique infinite path $\langle v_0, v_1, v_2 \dots \rangle$, where $\chi(v_i) = v_{i+1}$ if $v_i$ is in $V_{\text{Max}}$ and $\mu(v_i) = v_{i+1}$ if $v_i$ is in $V_{\text{Min}}$. This path, referred to as the play induced by the two strategies, will be denoted by $\text{Play}(\chi, \mu, v_0)$.

For all $v$ in $V$, we define $\text{Val}^*(v) = \min_{\mu \in \Pi_{\text{Min}}} \max_{\chi \in \Pi_{\text{Max}}} \mathcal{D}(\text{Play}(\chi, \mu, v))$, and $\text{Val}_*(v) = \max_{\chi \in \Pi_{\text{Max}}} \min_{\mu \in \Pi_{\text{Min}}} \mathcal{D}(\text{Play}(\chi, \mu, v))$. These will be known as the lower and upper values of $v$, respectively. It is always true that $\text{Val}_*(v) \leq \text{Val}^*(v)$. It is well known that for discounted games the two values are equal, a property known as determinacy.

**Theorem 1 ([17]).** *For every discounted game $G$ and every vertex $v \in V$, we have $\text{Val}_*(v) = \text{Val}^*(v)$.*

The value of the game starting at a vertex $v$, equal to both $\text{Val}_*(v)$ and $\text{Val}^*(v)$, is denoted by $\text{Val}(v)$. The computational task associated with discounted games is to compute Val($v$). Moreover, we want to find optimal strategies, i.e., a strategy $\chi$ that achieves the upper value and a strategy $\mu$ that achieves the lower value.

For convenience, we introduce the concept of a joint strategy $\sigma : V \to V$ that specifies moves for both players. The notation $\sigma \upharpoonright \text{Max}$ and $\sigma \upharpoonright \text{Min}$ will be used to refer to the individual strategies of Max and Min that constitute the joint strategy. For a vertex $v$, the function $\overline{\sigma}(v)$ gives the successor of $v$ not chosen by $\sigma$. The functions $r^\sigma$ and $r^{\overline{\sigma}}$ give the reward on the edge chosen by $\sigma$ and the reward on the edge not chosen by $\sigma$, respectively. The path denoted by $\text{Play}(\sigma, v)$ is equal to the path $\text{Play}(\sigma \upharpoonright \text{Max}, \sigma \upharpoonright \text{Min}, v)$. The joint strategy is optimal if both $\sigma \upharpoonright \text{Max}$ and $\sigma \upharpoonright \text{Min}$ are optimal. For a given joint strategy $\sigma$, the value of a vertex $v$ when $\sigma$ is played will be denoted by $\text{Val}^\sigma(v) = \mathcal{D}(\text{Play}(\sigma, v))$.

Given a joint strategy $\sigma$ and a vertex $v$, the *balance* of $v$ is the difference between the value of $v$ and the value of the play that starts at $v$, moves to $\overline{\sigma}(v)$ in the first step, and then follows $\sigma$,

$$\text{Bal}^\sigma(v) = \begin{cases} \text{Val}^\sigma(v) - (r^{\overline{\sigma}}(v) + \beta \cdot \text{Val}^\sigma(\overline{\sigma}(v))) & \text{if } v \in V_{\text{Max}}, \\ (r^{\overline{\sigma}}(v) + \beta \cdot \text{Val}^\sigma(\overline{\sigma}(v))) - \text{Val}^\sigma(v) & \text{if } v \in V_{\text{Min}}. \end{cases} \quad (1)$$

A vertex $v$ is said to be switchable under $\sigma$ if $\text{Bal}^\sigma(v) < 0$. If $\text{Bal}^\sigma(v) = 0$ for some vertex then that vertex is said to be indifferent. There is a simple characterisation of optimality in terms of switchable vertices.

**Theorem 2 ([17]).** *If no vertex is switchable in a joint strategy $\sigma$ then it is an optimal strategy for every choice of starting vertex.*

The two algorithms that we will present use only positional joint strategies. From now on, all joint strategies that we refer to can be assumed to be positional joint strategies. If a play begins at a vertex $v$ and follows a positional joint strategy $\sigma$ then the resulting infinite path can be represented by a simple path followed by an infinitely repeated cycle. Let $\text{Play}(\sigma, v) = \langle v_0, v_1, \ldots, v_{k-1}, \langle c_0, c_1, \ldots, c_{l-1} \rangle^\omega \rangle$. It is then easy to see that

$$\text{Val}^\sigma(v) = \sum_{i=0}^{k-1} \beta^i \cdot r^\sigma(v_i) + \sum_{i=0}^{l-1} \frac{\beta^{k+i}}{1-\beta^l} \cdot r^\sigma(c_i).$$

Therefore, the amount that the reward on the outgoing edge of a vertex $u$ contributes towards the value of $v$ can be defined as follows.

**Definition 3 (Contribution Coefficient).** *For vertices $v$ and $u$, and for a positional joint strategy $\sigma$, we define:*

$$\text{D}_\sigma^v(u) = \begin{cases} \beta^i & \text{if } u = v_i \text{ for some } 0 \leq i < k, \\ \frac{\beta^{k+i}}{1-\beta^l} & \text{if } u = c_i \text{ for some } 0 \leq i < l, \\ 0 & \text{otherwise.} \end{cases}$$

## 3 Lemke's Algorithm For Discounted Games

Lemke's algorithm is a classical algorithm for solving the linear complementarity problem [6]. We can apply Lemke's algorithm to a discounted game by utilising the reduction of Jurdziński and Savani [14], however this yields little insight into how the algorithm works on a discounted game. In this section we bypass the reduction, and give a description of Lemke's algorithm entirely in terms of discounted games.

Lemke's algorithm begins with the joint strategy $\sigma_0 = \rho$ that selects the right successor for every vertex in the game. This is actually a free choice since the left and right successors can be swapped to obtain an arbitrary starting strategy. The algorithm will then move through a sequence of strategies until it arrives at the optimal strategy. The algorithm will also construct a modified game for each strategy that it considers. The modified games will take the following form.

**Definition 4 (Modified Game For Lemke's Algorithm).** *For a real number $z$, we define the game $G_z$ to be the same as $G$ but with a modified left-edge reward function, denoted by $r_z^\lambda$, and defined, for every vertex $v$, by:*

$$r_z^\lambda(v) = \begin{cases} r^\lambda(v) - z & v \in V_{Max}, \\ r^\lambda(v) + z & v \in V_{Min}. \end{cases} \qquad (2)$$

For a modified game $G_z$, the function $r_z^\sigma$ will give the rewards on the edges chosen by $\sigma$. The notations $\text{Val}_z^\sigma$ and $\text{Bal}_z^\sigma$ will give the values the balances of the vertices in the game $G_z$, respectively. For every strategy $\sigma_i$ that is considered, the algorithm must choose an appropriate value $z_i$ so that $\sigma_i$ is optimal in $G_{z_i}$. Moreover, we want to choose the minimum value $z_i$ for which this property holds. The next proposition shows how to compute this for the initial strategy $\sigma_0$.

**Proposition 5.** *Let $z_0 = \max\{-\text{Bal}^{\sigma_0}(v) : v \in V\}$. The strategy $\sigma_0$ is optimal in $G_{z_0}$ and the vertex $v$ in $V$ that maximizes $-\text{Bal}^{\sigma_0}(v)$ is indifferent. Moreover, there is no value $y < z_0$ for which $\sigma_0$ is optimal in $G_y$.*

Proposition 5 gives an initial value for the parameter $z$. The principal idea behind the algorithm is to drive $z$ down from its initial value to 0, while maintaining optimality of the current strategy in $G_z$. Unfortunately, Proposition 5 implies that we cannot drive $z$ down further without losing the optimality of $\sigma_0$ in $G_z$. We do however know that there is some vertex $v$ that is indifferent under $\sigma_0$ in $G_{z_0}$. We define $\sigma_1 = \sigma_0[\overline{\sigma}_0(v)/v]$, i.e., $\sigma_1(u) = \overline{\sigma}_0(u)$ if $u = v$, and $\sigma_1(u) = \sigma_0(u)$ otherwise. The operation of modifying a strategy by changing the successor of a vertex $v$ will be referred to as switching $v$.

The value of no vertex changes when switching an indifferent vertex in a strategy. Since $\sigma_0$ was optimal in $G_{z_0}$ and $v$ was indifferent we therefore have that $\sigma_1$ is optimal in $G_{z_0}$. There is one important difference however, whereas $z$ could not be decreased without $\sigma_0$ losing its optimality, the parameter $z$ can be decreased further whilst maintaining the optimality of $\sigma_1$. The task now is to find $z_1$, the minimum value of $z$ for which $\sigma_1$ is still optimal.

At a high level, when the algorithm arrives at a strategy $\sigma_i$ its task is to find $z_i$, the minimum value of $z$ for which $\sigma_i$ is optimal in $G_z$. As we shall show, for this minimum value of $z$ there will always be at least one vertex that is indifferent under $\sigma_i$ played in $G_z$. The algorithm then switches this indifferent vertex to create $\sigma_{i+1}$ and the process is repeated. The remainder of this section is dedicated to showing how $z_i$ can be computed.

Each step begins with a strategy $\sigma_i$ and the value $z_{i-1}$, which was the minimum value of $z$ for which $\sigma_{i-1}$ was optimal in $G_z$. We now wish to know how much further $z$ can be decreased before $\sigma_i$ ceases to be optimal. From Theorem 2 we know that a strategy is optimal as long as no vertex is switchable and that a vertex is switchable only when it has a negative balance. It is for this reason that we want to know how the balance of each vertex changes as $z$ is decreased. In order to understand this, we must first know how the value of each vertex changes as $z$ is decreased. We will use the notation $\partial_{-z} \mathrm{Val}_z^\sigma(v)$ to denote the rate of change of the value of $v$ as $z$ decreases, i.e., $-\partial_z \mathrm{Val}_z^\sigma(v)$. This notation will be used frequently throughout the rest of the paper to denote the rate of change of various expressions. For a proposition $p$, we define $[p]$ to be equal to 1 if $p$ is true, and 0 otherwise. We can now give an explicit formula for $\partial_{-z} \mathrm{Val}_z^\sigma(v)$, which is based on the left edges that are passed through after visiting the vertex $v$ while playing the strategy $\sigma$, and the contribution coefficient of those edges to the value of $v$.

**Proposition 6.** *For a vertex $v$ and a joint strategy $\sigma$, let $L$ be the set of vertices for which $\sigma$ picks the left successor, $L = \{v \in V \;:\; \sigma(v) = \lambda(v)\}$. The rate of change of the value of $v$ is*

$$\partial_{-z} \mathrm{Val}_z^\sigma(v) = \sum_{u \in L} ([u \in V_{Max}] - [u \in V_{Min}]) \cdot \mathrm{D}_\sigma^v(u).$$

From equation (1) we know that the balance of a vertex is computed as a difference of the values of two vertices. We now show how the rate of change of the balance can be derived by substituting the rate of change of the values into equation (1).

**Proposition 7.** *For a vertex $v$ and a joint strategy $\sigma$ we have*

$$\partial_{-z} \mathrm{Bal}_z^\sigma(v) = \begin{cases} \partial_{-z}\mathrm{Val}_z^\sigma(v) - ([\overline{\sigma}(v) = \lambda(v)] + \beta \cdot \partial_{-z}\mathrm{Val}_z^\sigma(\overline{\sigma}(v))) & \text{if } v \in V_{Max}, \\ -[\overline{\sigma}(v) = \lambda(v)] + \beta \cdot \partial_{-z}\mathrm{Val}_z^\sigma(\overline{\sigma}(v)) - \partial_{-z}\mathrm{Val}_z^\sigma(v) & \text{if } v \in V_{Min}. \end{cases}$$

Now that we have an expression for the rate of change of the balance of a vertex, we can compute how far $z$ can be decreased from $z_{i-1}$ before some vertex gets a negative balance. For each vertex $v$, the expression $\mathrm{Bal}_{z_{i-1}}^{\sigma_i}(v)/\partial_{-z} \mathrm{Bal}_z^{\sigma_i}(v)$ gives the amount that $z$ can be decreased before $v$ gets a negative balance, and so the minimum over all these ratios gives the amount that $z$ can be decreased before some vertex gets a negative balance. It should also be clear that a vertex that achieves this minimum will be indifferent in the modified game when $z$ is decreased by this amount. We can also show that this is the minimum value of $z$ for which $\sigma_i$ is optimal in $G_z$.

**Proposition 8.** *Let a joint strategy $\sigma_i$ be optimal in the modified game $G_{z_{i-1}}$, and*
$$z_i = z_{i-1} - \min\{\frac{\mathrm{Bal}^\sigma_{z_{i-1}}(v)}{\partial_{-z}\mathrm{Bal}^\sigma_z(v)} \;:\; v \in V \text{ and } \partial_{-z}\mathrm{Bal}^\sigma_z(v) < 0\}. \qquad (3)$$
*Then strategy $\sigma$ is optimal in $G_{z_i}$, and it is not optimal in $G_x$ for all $x < z_i$.*

Until now, we have ignored the possibility of reaching a strategy $\sigma$ in which there is more than one indifferent vertex. In LCP algorithms this is known as a degenerate step. In this case, the task is to find a strategy in which every indifferent vertex $v$ satisfies $\partial_{-z}\mathrm{Bal}^\sigma_z(v) > 0$, so that $z$ can be decreased further. It is not difficult to prove that such a strategy can be reached by switching only the indifferent vertices. One method for degeneracy resolution is Bland's rule, which uses the least index method to break ties, and another is to use lexicographic perturbations; both methods are well-known, and are also used with the simplex method for linear programming [4].

---
**Algorithm 1** Lemke($G$)

$i := 0;\; \sigma_0 := \rho;\; z_0 := \max\{-\mathrm{Bal}^{\sigma_0}(v) \;:\; v \in V\}$
**while** $z_i > 0$ **do**
$\quad \sigma_{i+1} := \sigma_i[\overline{\sigma}_i(v)/v]$ for some vertex $v$ with $\mathrm{Bal}^{\sigma_i}_{z_i}(v) = 0$
$\quad z_{i+1} := z_i - \min\{\frac{\mathrm{Bal}^{\sigma_{i+1}}_{z_i}(v)}{\partial_{-z}\mathrm{Bal}^{\sigma_{i+1}}_z(v)} \;:\; v \in V \text{ and } \partial_{-z}\mathrm{Bal}^{\sigma_{i+1}}_z(v) < 0\}$
$\quad i := i + 1$
**end while**

---

Lemke's algorithm is shown as Algorithm 1. Since in each step we know that there is no value of $z < z_i$ for which $\sigma_i$ is optimal in $G_z$ and we decrease $z$ in every step it follows that we can never visit the same strategy twice without violating the condition that the current strategy should be optimal in the modified game. Therefore the algorithm must terminate after at most $2^{|V|}$ steps, which corresponds to the total number of joint strategies. The algorithm can only terminate when $z$ has reached 0, and $G_0$ is the same game as $G$. It follows that whatever strategy the algorithm terminates with must be optimal in the original game.

**Theorem 9.** *Algorithm 1 terminates, with a joint strategy $\sigma$ that is optimal for $G$ after at most $2^{|V|}$ iterations.*

Lemke's algorithm for LCPs allows a free choice of *covering vector*, and in our description we used a unit covering vector. This can be generalised by giving a positive covering value to every vertex. If each vertex $v$ has a covering value $d_v$ then the modification of the left edges in Definition 4 becomes:
$$r^\lambda_z(v) = \begin{cases} r^\lambda(v) - d_v \cdot z & v \in V_{\mathrm{Max}}, \\ r^\lambda(v) + d_v \cdot z & v \in V_{\mathrm{Min}}. \end{cases}$$
The algorithm can then easily be modified to account for this altered definition.

## 4 The Cottle-Dantzig Algorithm For Discounted Games

The principle idea behind the Cottle-Dantzig algorithm is to maintain a set of vertices whose balance is non-negative. The algorithm begins with an arbitrary strategy, and it goes through a series of major iterations, where in each iteration one vertex is brought into the set of vertices with non-negative balances, while maintaining the non-negative balances of the vertices that are already in that set. It is clear that if such a task can be accomplished, then the algorithm will terminate after $|V|$ major iterations.

We require a method for bringing some distinguished vertex $v$ into the set of vertices with a non-negative balance without the vertices currently in the set getting a negative balance in the process. To accomplish this we will modify the game by adding a bonus to the edge that the strategy currently chooses at $v$. We will then drive the bonus up from 0 while maintaining an optimal strategy for the modified game. Eventually the balance of $v$ will become 0 in the modified game, at which point the strategy at $v$ can be switched away from the edge with the bonus attached to it, and the bonus can be removed. We will prove that after this procedure $v$ will have a positive balance.

In this section we will override many of the notations that were used to describe Lemke's algorithm.

**Definition 10 (Modified Game For The Cottle-Dantzig Algorithm).**
*For a real number $w$, a joint strategy $\sigma$, and a distinguished vertex $v$, we define the game $G_w$ to be the same as $G$ but with a different reward on the edge chosen by $\sigma$ at $v$. If $\sigma$ chooses the left successor at $v$ then the left reward function is defined, for every $u$ in $V$, by:*

$$r_w^\lambda(u) = \begin{cases} r^\lambda(u) + w & \text{if } u = v \text{ and } u \in V_{Max}, \\ r^\lambda(u) - w & \text{if } u = v \text{ and } u \in V_{Min}, \\ r^\lambda(u) & \text{otherwise.} \end{cases}$$

*If $\sigma$ chooses the right successor at $v$ then $r^\rho$ modified in a similar manner.*

We begin the major iteration with a strategy $\sigma_0$, a value $w_0 = 0$, and a set of vertices with non-negative balances $P$. The task is to raise $w$ from 0 until $\text{Bal}_w^\sigma(v) = 0$, while maintaining the invariant that every vertex in $P$ has a non-negative balance. This can be accomplished using methods that are similar to those used in Lemke's algorithm. For every vertex in $P$ we must compute how the balance of that vertex changes as $w$ is increased. The following propositions are analogues of Propositions 6, 7, and 8.

**Proposition 11.** *Consider a vertex $u$ and a joint strategy $\sigma$. Suppose that $v$ is the distinguished vertex. The rate of change $\partial_w \text{Val}_w^\sigma(u)$ is $\text{D}_\sigma^u(v)$.*

**Proposition 12.** *Consider a vertex $u$ and a joint strategy $\sigma$ in the game $G_w$. The rate of change $\partial_w \text{Bal}_w^\sigma(u)$ is:*

$$\partial_w \text{Bal}_w^\sigma(u) = \begin{cases} \partial_w \text{Val}_w^\sigma(u) - \beta \cdot \partial_w \text{Val}_w^\sigma(\overline{\sigma}(u)) & \text{if } u \in V_{Max}, \\ \beta \cdot \partial_w \text{Val}_w^\sigma(\overline{\sigma}(u)) - \partial_w \text{Val}_w^\sigma(u) & \text{if } u \in V_{Min}. \end{cases}$$

**Algorithm 2** Cottle-Dantzig($G$, $\sigma$)

$P := \emptyset$
**while** $P \neq V$ **do**
  $i := 0$; $w_0 := 0$; $v :=$ Some vertex in $V \setminus P$
  **while** $\text{Bal}^\sigma_{w_i}(v) < 0$ **do**
    $w_{i+1} := w_i + \min\{-\frac{\text{Bal}^\sigma_{w_i}(u)}{\partial_w \text{Bal}^\sigma_w(u)} \ : \ u \in P \cup \{v\}$ and $\partial_w \text{Bal}^\sigma_w(u) < 0\}$
    $\sigma := \sigma[\overline{\sigma}(u)/u]$ for some vertex $u$ with $\text{Bal}^\sigma_{w_i}(v) = 0$
    $i := i + 1$
  **end while**
  $\sigma := \sigma[\overline{\sigma}(v)/v]$; $P := P \cup \{v\}$
**end while**

**Proposition 13.** *Consider a modified game $G_w$, a joint strategy $\sigma$, and a set of vertices $P$ which must not have negative balances. Let*

$$y = w + \min\{-\frac{\text{Bal}^\sigma_w(u)}{\partial_w \text{Bal}^\sigma_w(u)} \ : \ u \in P \cup \{v\} \text{ and } \partial_w \text{Bal}^\sigma_w(u) < 0\}.$$

*No vertex in $P$ has a negative balance in $G_y$. Moreover, one vertex in $P \cup \{v\}$ is indifferent, and for all values $x > y$ that vertex has a negative balance in $G_x$.*

The process of raising $w$ up from 0 until the balance of $v$ is 0 in the modified game is the same as the process of decreasing $z$ in Lemke's algorithm, only using the different definitions from Propositions 11, 12, and 13. Once the balance of $v$ has reached 0 we can stop increasing $w$. Since $v$ is now indifferent we can switch it away from the edge that has the bonus attached to it. Once this has been done, the values of all vertices are no longer affected by $w$, since the edge to which it is attached is no longer chosen by the current strategy. Therefore we can remove the bonus and recover the original game. The major iteration then terminates with a strategy in which every vertex in $P \cup \{v\}$ has a non-negative balance, and the next major iteration can begin.

**Theorem 14.** *Algorithm 2 terminates, with the optimal joint strategy, after at most $2^{|V|}$ iterations.*

## 5 Exponential Lower Bounds

We show that both Lemke's and the Cottle-Dantzig algorithms take exponentially many steps on the family of games shown in Figure 1. Max vertices are depicted as squares and Min vertices are depicted as circles. For every vertex, we define the right successor to be the vertex with the same owner as the vertex itself, and the left successor to be the vertex that belongs to the other player. Recall that the initial strategy for Lemke's algorithm is the one that chooses the right successor for every vertex. When speaking about vertices in the game we often refer to either the leftmost or the rightmost vertex with a certain property.

In this context, the vertex being referred to is the one that is furthest to the right or to the left in Figure 1.

For ease of exposition, we will describe the steps of the algorithm as if the discount factor was 1. Although this is forbidden by the definition of a discounted game, since the game contains one cycle, whose value is zero, the value of every vertex under every strategy will be finite. As long as the discount factor is chosen sufficiently close to 1, the algorithm will behave as we describe.

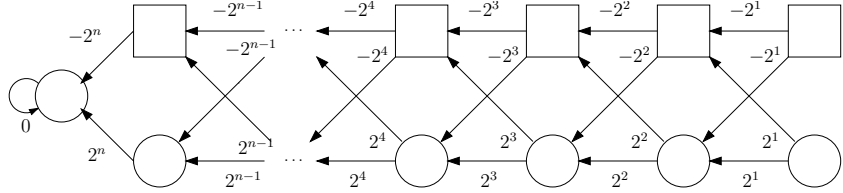

**Fig. 1.** The game $\mathcal{G}_n$.

Note that the game graph is symmetric with respect to the line that separates the vertices of the two players. We frequently refer to a vertex and the vertex that it is opposite to, and hence we introduce the concept of vertex reflections. For a vertex $v$ that is not the sink, we write $\overline{v}$ to denote the reflection of $v$, that is the vertex belonging to the other player that is shown directly opposite $v$ in Figure 1. We say that a joint strategy $\sigma$ is symmetric if for all vertices $v$, the strategy $\sigma$ chooses the right successor of $v$ if and only if it chooses the right successor of $\overline{v}$. The initial strategy for Lemke's algorithm is a symmetric strategy. Lemke's algorithm always switches $v$ directly before or after $\overline{v}$ and so it can be seen as traversing through symmetric strategies.

Before discussing the modified games that Lemke's algorithm constructs, we give a simple characterisation of when a vertex is switchable in the original game.

**Proposition 15.** *If $\sigma$ is a symmetric joint strategy, then a vertex $v$ is switchable if and only if the path from $v$ has an even number of left edges.*

We now use this characterisation to give a simple formula for $\partial_{-z} \operatorname{Bal}^\sigma_z(v)$ for every vertex $v$ under every symmetric joint strategy $\sigma$.

**Proposition 16.** *If $\sigma$ is a symmetric joint strategy, then $\partial_{-z} \operatorname{Bal}^\sigma_z(v)$, the rate of change of the balance of a vertex $v$, is 1 if $v$ is switchable, and $-1$ otherwise.*

Together, Propositions 15 and 16 imply that the parameter $z$ can be set to the largest balance of a switchable vertex. We show that the largest balance will always belong to the rightmost switchable vertex.

**Proposition 17.** *Let $\sigma$ be a symmetric joint strategy, $v$ be the rightmost switchable Max vertex, and $z = -\operatorname{Bal}^\sigma(v)$. Then no vertex in $G_z$ is switchable, both $v$*

*and the reflection of v are indifferent, and for every real number $y < z$, there is a switchable vertex in $G_y$.*

Proposition 17 implies that whenever Lemke's algorithm is considering a symmetric joint strategy, it must choose a $z$ so that the rightmost switchable vertex is indifferent. We show that this leads to an exponential number of switches.

**Theorem 18.** *Lemke's algorithm performs $2^{n+1} - 2$ iterations on the game $\mathcal{G}_n$.*

The Cottle-Dantzig algorithm is sensitive to the order in which to bring the vertices into the non-negative set. We prove that there is an order that causes exponential-time behaviour for this algorithm. The sequence of strategies is similar to the sequence that Lemke's algorithm follows.

**Theorem 19.** *Consider an order in which all Min vertices precede Max vertices, and Max vertices are ordered from right to left. The Cottle-Dantzig algorithm performs $2^{n+1} - 1$ iterations.*

We have shown that both algorithms can take an exponential number of steps on the discounted game $\mathcal{G}_n$. We argue that this also implies an exponential lower bound for parity and average-reward games. From the game $\mathcal{G}_n$ we can obtain a parity game by replacing the reward $\pm 2^c$ with the priority $c$. The standard reductions [16, 21, 13] convert this parity game into average-reward and discounted games where priority $c$ is replaced with reward $(-n)^c$, and the discount factor is chosen to be very close to 1. All arguments used to prove Theorems 18 and 19 continue to hold if rewards of magnitude $2^c$ are replaced with rewards of magnitude $n^c$, which implies the exponential lower bounds also hold for parity games and average-reward games.

## 6  Future Work

Our adaptation of Lemke's algorithm for solving discounted games corresponds to its implementation in which the unit covering vector is used [6], and our lower bounds are specific to this choice. Similarly our lower bounds for the Cottle-Dantzig algorithm require a specific choice of ordering over the vertices. Randomizing these choices may exhibit better performance and should be considered.

Adler and Megiddo [1] studied the performance of Lemke's algorithm for the LCPs arising from linear programming problems. They showed that, for randomly chosen linear programs and a carefully selected covering vector, the expected number of pivots performed by the algorithm is quadratic. A similar analysis for randomly chosen discounted games should be considered.

# A Proofs for Section 3

## A.1 Proof of Proposition 5

*Proof.* If there is no vertex $v$ that satisfies $\text{Bal}^{\sigma_0}(v) < 0$ then the current strategy is optimal by Theorem 2 and there is no need to modify the game. Otherwise, the definition of the modified game states that for every Max vertex, the weight on the edge not chosen by $\sigma_0$ is decreased by $z_0$. Similarly for every Min vertex, the weight on the edge not chosen by $\sigma_0$ is increased by $z_0$. Since none of the modified weights are on edges chosen by $\sigma_0$ every vertex has the same value in $G_{z_0}$ as it does in $G$ when $\sigma_0$ is played, and only the balance of the vertices will change. These three observations imply that the balance of all vertices, as given by equation (1), is increased by $z_0$. Since $z_0$ was chosen to be equal in magnitude to the largest negative balance, the balance of every vertex when $\sigma_0$ is played in $G_{z_0}$ must be non-negative. Therefore $\sigma_0$ is optimal in $G_{z_0}$ by Theorem 2. It is also easy to see that the vertex with the largest negative balance is indifferent and that this vertex would have a negative balance for all games modified by a value smaller than $z_0$. □

## A.2 Proof of Proposition 6

*Proof.* From equation (2) we find that for a vertex $v$,

$$r^{\sigma}_{z-c}(v) = \begin{cases} r^{\sigma}_z(v) + c & \text{if } v \in V_{\text{Max}} \text{ and } \sigma(v) = \lambda(v), \\ r^{\sigma}_z(v) - c & \text{if } v \in V_{\text{Min}} \text{ and } \sigma(v) = \lambda(v), \\ r^{\sigma}_z(v) & \text{otherwise.} \end{cases}$$

Therefore, the change in $\text{Val}^{\sigma}(v)$ is equal to:

$$\text{Val}^{\sigma}_{z-c}(v) - \text{Val}^{\sigma}_z(v)$$
$$= \sum_{i=0}^{k-1} \beta^i \cdot r^{\sigma}_{z-c}(v_i) + \sum_{i=0}^{l-1} \frac{\beta^{k+i}}{1-\beta^l} \cdot r^{\sigma}_{z-c}(c_i) - \text{Val}^{\sigma}_z(v)$$
$$= \sum_{x \in L} ([x \in V_{\text{Max}}] \cdot D^v_{\sigma}(x) - [x \in V_{\text{Min}}] \cdot D^v_{\sigma}(x)) \cdot c + \text{Val}^{\sigma}_z(v) - \text{Val}^{\sigma}_z(v)$$
$$= \sum_{x \in L} ([x \in V_{\text{Max}}] \cdot D^v_{\sigma}(x) - [x \in V_{\text{Min}}] \cdot D^v_{\sigma}(x)) \cdot c \ .$$

□

## A.3 Proof of Proposition 7

*Proof.* The formula is obtained by substituting $\partial_{-z} \text{Val}^{\sigma}_z$ for $\text{Val}^{\sigma}$ in Definition 1. Additional care must be taken if $\sigma$ chooses the right successor of $v$. In this case, the left edge of $v$ will also be decreased as $z$ is decreased, and this is not captured by $\partial_{-z} \text{Val}^{\sigma}_z(\overline{\sigma}(v))$. This can be corrected for, however, by substituting $[\overline{\sigma}(v) = \lambda(v)]$ for $r^{\overline{\sigma}}(v)$ in Definition 1. □

## A.4 Proof of Proposition 8

*Proof.* All vertices with $\partial_{-z} \operatorname{Bal}_z^{\sigma_i} \geq 0$ can be ignored since their balances will not decrease as $z$ is decreased from $z_{i-1}$. For every vertex $v$ with $\partial_{-z} \operatorname{Bal}_z^{\sigma_i}(v) < 0$, the ratio $(\operatorname{Bal}_z^{\sigma_i}(v))/(\partial_{-z} \operatorname{Bal}_z^{\sigma_i}(v))$ gives the largest amount that $z$ can be decreased by before $\operatorname{Bal}_z^{\sigma_i}(v)$ becomes equal to zero. By choosing $z_i$ to be the minimum over these ratios we ensure that the balance of all vertices remains non-negative while the vertex with the minimum ratio becomes indifferent. For all values $x < z_i$ the vertex with the minimum ratio will have a negative balance, which implies $\sigma_i$ will not be optimal in $G_x$ by Theorem 2. □

## A.5 Proof of Theorem 9

*Proof.* In each step we have a game $G_{z_i}$ in which the current strategy $\sigma_i$ is optimal. We also know that there is a unique vertex $v$ that is indifferent, and that $\partial_{-z} \operatorname{Bal}_z^{\sigma_i}(v) < 0$, which implies that $z$ cannot be decreased further without the balance of $v$ becoming negative. The vertex $v$ is then switched in $\sigma_i$, giving $\sigma_{i+1}$. Note that by Proposition 7, the balance of $v$ is computed as the difference between the successor chosen by the current strategy and the alternate successor. Switching $v$ causes this difference to be reversed and therefore, we have:

$$\partial_{-z} \operatorname{Bal}_z^{\sigma_i}(v) = -\partial_{-z} \operatorname{Bal}_z^{\sigma_{i+1}}(v).$$

It follows that as $z$ is decreased, $\operatorname{Bal}^{\sigma_{i+1}}(v)$ will increase. Since $v$ was the unique indifferent vertex, $z$ can be decreased by a strictly positive amount while maintaining optimality of $\sigma_{i+1}$.

Proposition 8 implies that $\sigma_i$ is not optimal in $G_x$ for all $x < z_i$. Since $z_{i+j} < z_i$ for all $j > 0$ it follows that $\sigma_i$ will never again be an optimal strategy for the modified game. The algorithm always maintains optimality of the current strategy in the modified game, and so it can never revisit $\sigma_i$. There are only $2^{|V|}$ possible strategies and therefore, the algorithm must terminate after at most $2^{|V|}$ iterations. The algorithm must terminate at the optimal strategy since optimality of the current strategy in the modified game is maintained in each iteration and it cannot terminate until $z$ reaches zero, and $G_0$ is the same as the original game. □

## B Proofs for Section 4

### B.1 Proof of Proposition 11

*Proof.* The proof is very similar to the proof of Proposition 6. Since $\operatorname{D}_\sigma^u(v)$ gives the precise contribution of the weight on the edge of $v$ to the value of $u$ and the outgoing edge of $v$ is the only edge that is modified in $G_w$ it is clear that the value of $u$ will increase in proportion to $\operatorname{D}_\sigma^u(v)$. □

## B.2 Proof of Proposition 12

*Proof.* The proof is very similar to the proof of Proposition 7. We obtain the expression by substituting $\partial_w \text{Val}^\sigma_w(v)$ for $\text{Val}^\sigma(v)$ in Definition 1. In contrast to the expression used for Lemke's algorithm, we do not need to account for the edge not chosen by $\sigma$ at $u$. This is because only one edge is modified in $G_w$, and it is chosen by all strategies during the current major iteration. □

## B.3 Proof of Theorem 14

*Proof.* As with Lemke's algorithm, we assume for now that there are no degenerate steps during the execution of the algorithm, as these can be resolved using the same rules that were outlined for Lemke's algorithm. Using an identical argument as the one given in Theorem 9, it can be shown that during each major iteration the algorithm can pass through at most $2^{|P|}$ different strategies. Since the set $P$ is equal to $V$ after the final major iteration, and all vertices in $P$ may never have a negative balance, all vertices will not be switchable in the final strategy, and it must therefore be optimal by Theorem 2. To see that the algorithm terminates after considering at most $2^{|P|}$ strategies, note that each major iteration passes through $2^{|P|}$ strategies. The maximum number of strategies visited must therefore be:

$$\sum_{i=0}^{|V|-1} 2^i = 2^{|V|}.$$

□

## C Proofs for Section 5

### C.1 Proof of Proposition 15

*Proof.* If the vertex $v$ belongs to player Max and the path from $v$ uses an even number of left edges then the final edge used before reaching the loop will have weight $-2^n$. By symmetry, the path starting at the alternate successor of $v$ will pass through the weight $2^n$. If we ignored the other weights on the two paths then the balance of $v$ would be $2 \times -2^n$ and $v$ would be switchable. It can easily be verified that no matter which other edges the path passes through, the sum of the weights of those edges can never reach $2 \times 2^n$. This means that the balance of $v$ cannot be brought above 0 and the vertex will therefore be switchable. Conversely, if the path uses an odd number of left edges then the difference of the final edge used will be $2 \times 2^n$ and there is no path that has enough weight to bring the balance of the vertex below zero, implying that the vertex is not switchable. The arguments work symmetrically for vertices owned by player Min. □

## C.2 Proof of Proposition 16

*Proof.* We prove the proposition for the case when $v$ is switchable. The proof for the case where $v$ is not switchable very similar. Since $v$ is switchable, the path from $v$ that follows $\sigma$ contains an even number of left edges. Since the owner of vertex through which the path passes changes only when a left edge is taken it follows that precisely half of the left edges originate from vertices belonging to either player. If $X$ is the set of vertices for which the path from $v$ uses a left edge, then by Proposition 6 the rate of change of $v$ is:

$$\partial_{-z} \operatorname{Val}_z^\sigma(v) = \sum_{x \in X} ([x \in V_{\text{Max}}] \cdot \mathrm{D}^v(x) - [x \in V_{\text{Min}}] \cdot \mathrm{D}^v(x))$$
$$= \sum_{x \in X} ([x \in V_{\text{Max}}] - [x \in V_{\text{Min}}]) = 0.$$

If $\sigma$ chooses the left successor of $v$ then the path starting from $\sigma(v)$ that follows $\sigma$ contains an odd number of left edges, and by symmetry the path that follows $\sigma$ from the alternate successor of $v$ has an odd number of left edges. Alternatively, if $\sigma$ chooses the right successor of $v$ then the path that follows $\sigma$ from $\sigma(v)$ contains an even number of left edges and so does the path that follows $\sigma$ from the alternate successor of $v$. Either way, the path starting at $v$ that moves to $\overline{\sigma}(v)$ and then follows $\sigma$ contains an odd number of left edges. Moreover the final left edge on that path emanates from a vertex owned by the same player that owns $v$. Therefore, by Proposition 6

$$[\overline{\sigma}(v) = \lambda(v)] + \beta \cdot \partial_{-z} \operatorname{Val}_z^\sigma(\overline{\sigma}(v)) = \begin{cases} 1 & \text{if } v \in V_{\text{Max}}, \\ -1 & \text{if } v \in V_{\text{Min}}. \end{cases}$$

The proposition can now be proved by substituting the values into Proposition 7. □

## C.3 Proof of Proposition 17

*Proof.* Let $\{v_0, v_1, \ldots, v_k\}$ be the set of switchable vertices belonging to player Max ordered left to right. By Proposition 15 we know that there are an even number of left edges on the path that follows $\sigma$ from each of these vertices. From a vertex $v_i$, the path either takes a right edge to $v_{i-1}$ or takes a left edge to a Min vertex, followed by a possibly empty sequence of right edges passing through Min vertices followed by a second left edge back to a Max vertex $u$. Since the path from $v_i$ passes through an even number of left edges the path from all of the odd vertices passed through must have an odd number of left edges indicating that they are all unprofitable and by symmetry all of the Max vertices between $v_i$ and $u$ are also unprofitable. It follows that $u$ is equal to $v_{i-1}$ and that the path from $v_i$ passes through every vertex $v_j$ with $j < i$.

Since both of the outgoing edges of $v_i$ have the same weight the balance at $v_i$ is:

$$\operatorname{Bal}^\sigma(v_i) = \operatorname{Val}^\sigma(\sigma(v_i)) - \operatorname{Val}^\sigma(\overline{\sigma}(v_i)).$$

Since the path leaving $v_{i+1}$ that follows $\sigma$ passes through $v_i$, by symmetry the path leaving the alternate edge of $v_{i+1}$ must pass through the reflection of $v_{i+1}$ and then $\overline{\sigma}(v_i)$. Let $c$ and $d$ denote the sum of the weights along the path from $v_{i+1}$ to $v_i$ and from the alternate edge of $v_{i+1}$ to the reflection of $v_i$. The balance at $v_{i+1}$ is then:

$$\mathrm{Bal}^\sigma(v_{i+1}) = (\mathrm{Val}^\sigma(\sigma_(v_i)) + r_\sigma(v_i) + c) - (\mathrm{Val}^\sigma(\overline{\sigma}(v_i)) + r_\sigma(\overline{v_i}) + d)$$
$$= \mathrm{Bal}^\sigma(v_i) + (r_\sigma(v_i) + c) - (r_\sigma(\overline{v_i}) + d).$$

Note that both $r_\sigma(v_i)$ and $-r_\sigma(\overline{v_i})$ are both negative and that $c - d$ cannot possibly cancel them out. Therefore:

$$\mathrm{Bal}^\sigma(v_i) > \mathrm{Bal}^\sigma(v_{i+1}).$$

Now, if $z$ is equal to $-\mathrm{Bal}^\sigma(v_k)$, since all other switchable vertices have balances with smaller magnitude none of them will be switchable in $G_z$ and $v_k$ will be indifferent. For any real parameter $y$ less than $z$ the vertex $v_k$ would be switchable in $G_y$. □

### C.4 Proof of Theorem 18

*Proof.* The claim will be proved by induction. In $G_1$ there are only two vertices with more than one successor and they are reflections of each other. Initially they are both profitable, since their paths use zero left edges and in the first and second iterations they switch, arriving at the optimal strategy in two iterations.

The game $G_i$ can be broken down into the leftmost pair of vertices and a game $G_{i-1}$ which encompasses the vertices to the right of this pair. Since Lemke's algorithm works from the right an optimal strategy for the game $G_{i-1}$ must be computed before the algorithm will touch the leftmost pair. Once the algorithm has arrived at this strategy the leftmost pair will both be switched, one after the other. Suppose that while solving the sub-game $G_{i-1}$, the algorithm passed through the sequence of strategies $\langle \sigma_0, \sigma_1, \ldots, \sigma_k \rangle$. Let $\sigma'_i$ be the strategy $\sigma_i$ with the left pair switched. Note that due to Proposition 15 a vertex pair is switchable in $\sigma_i$ if and only if it is not switchable in $\sigma'_i$. Suppose that $v$ and $\overline{v}$ are the rightmost switchable pair of vertices in $\sigma'_i$. After switching the pair, the algorithm moves to a strategy $\tau$ in which $v$ and $\overline{v}$ are not switchable and everything to the right of them is switchable. If the leftmost pair were not switched then the situation would be precisely reversed with $v$ and $\overline{v}$ being the rightmost switchable pair. It follows that $\tau = \sigma'_{i-1}$ and therefore after switching the leftmost pair the algorithm will then proceed to pass through all of the strategies that it has visited up to that point in reverse order. From this we obtain the recursion:

$$T(1) = 2,$$
$$T(n) = T(n-1) + 2 + T(n-1).$$

It can easily be verified that $T(n) = 2^{n+1} - 2$. □

## D  Exponential Lower Bound For The Cottle-Dantzig Algorithm (Proof of Theorem 19)

We show that the example that was used to show an exponential lower bound for Lemke's algorithm can also be used to show an exponential lower bound for the Cottle-Dantzig algorithm. The initial strategy will be the one that selects the right successor for every Max vertex and the left successor for every Min vertex. Note that for this strategy every Min vertex has a positive balance and every Max vertex has a negative balance. Recall that the Cottle-Dantzig algorithm allows a free choice for the order in which vertices are brought into the non-negative set. We use the following order: all of the Min vertices will be brought in first, followed by the Max vertices, which will be brought in from right to left. Since the balance of all Min vertices is positive, bringing them into the non-negative set is a trivial operation that does not require modification of the initial strategy. The major iterations that follow will be numbered 1 to $n$, and it is our goal to show that the Cottle-Dantzig algorithm will take $2^i - 1$ steps in major iteration $i$.

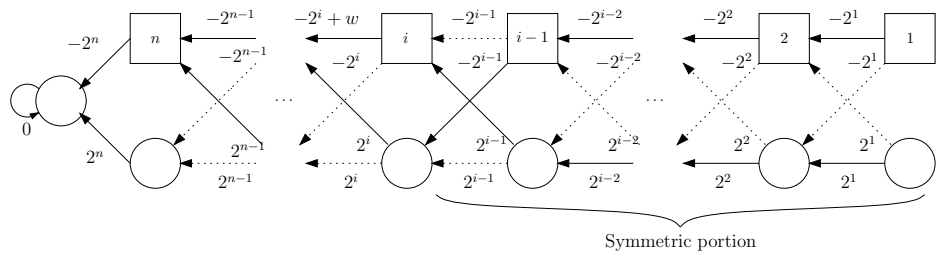

**Fig. 2.** The state of the algorithm at the start of major iteration $i$.

The situation at the start of major iteration $i$ is depicted in Figure 2. The edges chosen by the current strategy are depicted as solid edges, while dotted edges show the alternate edges. The indices on the Max vertices show in which iteration they are designated to enter the non-negative set. For convenience, we use these indices to identify both the vertices themselves and the pair consisting of the vertex with index $i$ and its reflection. It can easily be seen that in major iteration $i$ the strategy may only change at vertices that are to the right of the vertex pair with index $i$, since no other vertices can reach the Max vertex with index $i$. Unlike Lemke's algorithm, the overall strategy will not be symmetric until the optimal strategy is found, but after major iteration $i$ the strategy will be symmetric on the $i$ rightmost vertex pairs. This is shown as the symmetric portion in Figure 2.

We begin by giving a characterisation of $\partial \operatorname{Bal}_z^\sigma$ for the vertices in the symmetric portion.

**Proposition 20.** *Suppose that the algorithm is in the major iteration $i$ at a strategy $\sigma$, and that $v$ is a vertex in the symmetric portion of $\sigma$. Let $\pi$ be the path from $v$ to the vertex with index $i$ or its reflection, according to $\sigma$. Then we have:*

$$\partial_z \operatorname{Bal}_z^\sigma(v) = \begin{cases} -1 & \text{If } \pi \text{ contains an odd number of left edges,} \\ 1 & \text{otherwise.} \end{cases}$$

*Proof.* If $v$ is a Max vertex and the path from $v$ to vertex pair with index $i$ takes an odd number of left edges then the path will arrive at the reflection of Max vertex with index $i$. It follows that $\partial_z \operatorname{Val}_z^\sigma(v)$ is equal to 0, since the path does not pass through the Max vertex with index $i$. Now consider the alternate successor of $v$, which we call $u = \bar\sigma(v)$. Since the strategy is symmetric until it reaches the vertex pair with index $i$, the path from $u$ must lead to the Max vertex with index $i$ and so $\partial_z \operatorname{Val}_z^\sigma(u)$ is equal to 1. Substituting these into the definition of $\partial_z \operatorname{Bal}_z^\sigma(v)$ gives:

$$\partial_z \operatorname{Bal}_z^\sigma(v) = \partial_z \operatorname{Val}_z^\sigma(v) - \beta \cdot \partial_z \operatorname{Val}_z^\sigma(u) = 0 - 1 = -1.$$

If the path from $v$ uses an even number of left edges then the above is reversed: the path from $v$ passes through the Max vertex with index $i$ and the path from $u$ does not. With the same reasoning we can conclude that $\partial_z \operatorname{Bal}_z^\sigma(v) = 1$. The proof for vertices belonging to player Min is entirely symmetrical. □

Proposition 20 implies that for the initial strategy $\sigma$ in major iteration $i$, every vertex $v$ to the right of the vertex pair with index $i$ has $\partial_z \operatorname{Bal}_z^\sigma(v)$ equal to $-1$. It follows that as $w$ is increased, the first vertex to switch will be the one with the smallest balance. For the initial strategy this will be the rightmost vertex pair.

**Proposition 21.** *Consider the first strategy in major iteration $i$. For every vertex $v$ with index $j$ which is smaller than $i$, the balance of $v$ will be $2 \times (2^i - \sum_{k=j+1}^{i-1} 2^k)$*

*Proof.* It can be seen in Figure 2 that Max vertex with index $i-1$ has balance:

$$2^i - 2^{i-1} - (-2^i - 2^{i-1}) = 2 \times 2^i$$

The Min vertex with index $i-1$ has the same balance. Now, for every vertex to the right of the Max vertex $v$ with index $i-1$, the balance can be computed by considering two paths: the path that follows $\sigma$ from $v$ to the sink and the path that uses the alternate edge of $v$ and then follows $\sigma$ to the sink. These two paths join at the Max vertex with index $i+1$, and so the weights after this point are irrelevant. The two weights on the outgoing edges are also irrelevant since they are identical and one will be subtracted from the other. If $v$ is the Max vertex with index $j$ then its balance is:

$$2^i - 2^{i-1} - \cdots - 2^j - (-2^i + 2^{i-1} + \cdots + 2^{j+1} - 2^j)$$
$$= 2 \times (2^i - \sum_{k=j+1}^{i-1} 2^k)$$

A symmetric proof can be used to show that the Min vertices have identical balances to their reflections. □

Our goal is to show that major iteration $i$ will take $2^i - 2$ steps. For this purpose we define, for $1 \leq j < i$, the quantity $k_j$ to be equal to the number of strategies that the Cottle-Dantzig algorithm passes through before the vertex with index $j$ has balance 0 for the first time. Furthermore, we define $w_j$ to be the value of the parameter $w$ the first time the vertex with index $j$ has balance zero. Note that Proposition 20 implies that the first value of $w$ chosen by the algorithm will be the minimum over the balances of the vertices in the symmetric portion. Proposition 21 implies that this will be the rightmost pair of vertices, whose indices are 1. Therefore we set $k_1 = 0$ and $w_1 = 2 \times (2^i - \sum_{k=2}^{i-1} 2^k)$. The rightmost pair of vertices will be indifferent in the game $G_{w_1}$. We use these as the base case for the following inductive proposition.

**Proposition 22.** *Suppose that the algorithm is in major iteration $i$ and that is has passed through $k_j$ strategies, arriving at the first strategy in which the vertices with index $j$ have balance 0. All of the following hold:*

- *The algorithm will pass through $k_j + 2$ further strategies before the vertex with index $j+1$ has balance 0 for the first time. That is $k_{j+1} = 2k_j + 2$.*
- *The value of $w_{j+1}$, which is equal to the value of the parameter $w$ when the vertex with index $j+1$ has balance 0 for the first time, will be $w_{j+1} = 2 \times (2^i - \sum_{k=j+2}^{i-1} 2^k)$.*
- *No vertex with index higher than $j+1$ will be switched before the vertex with index $j+1$ has balance 0.*

*Proof.* Let $\langle \sigma_0, \sigma_1, \ldots \sigma_{k_j} \rangle$ be the sequence of strategies that the algorithm considered before the vertex with index $j$ had balance 0 for the first time. For a strategy $\sigma_l$ in this sequence we define the strategy $\sigma'_l$ as follows, for every vertex $v$:

$$\sigma'_l(v) = \begin{cases} \overline{\sigma}_l(v) & \text{if the index of } v \text{ is } j, \\ \sigma_l(v) & \text{otherwise.} \end{cases}$$

In other words, $\sigma'_l$ is $\sigma_l$ in which the vertices with index $j$ have been switched. Since both vertices with index $j$ are indifferent in $\sigma_{k_j}$ the algorithm will spend two iterations switching them, one after the other. Note that the algorithm has now arrived at $\sigma'_{k_j}$. Proposition 20 implies that for every $l$ in the range $1 \leq l \leq k_j$ and every vertex $v$ with index smaller than $j$,

$$\partial_z \operatorname{Bal}_z^{\sigma_l}(v) = -\partial_z \operatorname{Bal}_z^{\sigma'_l}(v).$$

This is because after switching the vertex pair with index $j$ every vertex with index less than $j$ sees an extra left edge on its path to the cycle. So, as $w$ is increased the balance of every vertex $\sigma'_l$ will move in a direction that is opposite to the way that it moved in $\sigma_l$. Under the assumption that no vertex with index higher than $j$ becomes indifferent, it can easily be verified that if $w$ is raised by $w_j - w_1$ the Cottle-Dantzig algorithm will pass through the sequence of strategies $\langle \sigma'_{k_j}, \sigma'_{k_j-1}, \ldots \sigma'_0 \rangle$ which together with the two iterations spent switching the vertices with index $j$ implies that the algorithm will pass through $2k_j + 2$ further strategies.

We now prove the assumption that the algorithm can raise $w$ to $w_j + (w_j - w_1)$ while not touching the vertices with indices higher than $j$. For a vertex $v$ with index $m$, which is higher than $j$, Proposition 21 implies that the balance of $v$ at the start of major iteration $i$ was:

$$2 \times (2^i - \sum_{k=m+1}^{i-1} 2^k).$$

Furthermore, the inductive hypothesis guarantees that the strategy has not been changed from the initial strategy on all vertices with index higher than $j$. Since in the initial strategy $\partial_z \operatorname{Bal}_z^{\sigma_0}(v) = -1$ it follows that $\partial_z \operatorname{Bal}_z^{\sigma_l}(v) = -1$ and $\partial_z \operatorname{Bal}_z^{\sigma'_l}(v) = -1$ for all $l$ in the range $1 \leq l \leq k_j$. Hence, the balance of $v$ has been decreasing continuously since the start of major iteration $i$. Therefore, in iteration $k_j$, when the vertex with index $j$ had balance 0 for the first time, the balance of $v$ was:

$$2 \times (2^i - \sum_{k=m+1}^{i-1} 2^k) - w_j$$
$$= 2 \times (2^i - \sum_{k=m+1}^{i-1} 2^k) - (2 \times (2^i - \sum_{k=j+1}^{i-1} 2^k))$$
$$= 2 \times \sum_{k=j+1}^{m} 2^k.$$

From this we can conclude that the balance of every vertex with index higher than $j$ is at least $2 \times 2^{j+1}$. We intend to raise $w$ by an amount equal to $w_j - w_1$, this is equal to:

$$2 \times (2^i - \sum_{k=j+1}^{i-1} 2^k) - (2 \times (2^i - \sum_{k=2}^{i-1} 2^k))$$
$$= 2 \times \sum_{k=2}^{j} 2^j < 2 \times (2^{j+1} - 1)$$

Since this value is smaller than $2 \times 2^{j+1}$ no vertex with index higher than $j$ will become negative while $w$ is being raised.

So far we have proved that the algorithm will pass through $k_j + 2$ further strategies without changing the strategy at the vertices with indices higher than $j$. We now prove that in iteration $2k_j + 2$ the vertex pair with index $j + 1$ will be indifferent. Note that in $\sigma'_0$ the path from every vertex with index less than or equal to $j$ uses precisely two left edges: one which belongs to a vertex with index $j$ and one that belongs to a vertex with index $i-1$. From this we can conclude, by Proposition 20, that $\partial_z \operatorname{Bal}_z^{\sigma_0'}(v) = 1$ for every vertex $v$ with index smaller than or equal to $j$. This implies that once the algorithm has reached $\sigma_0'$ it can continue to raise $w$ until the balance of some vertex with index greater than $j$ becomes 0. We have already shown that the balance of every vertex with index higher than $j$ has decreased in every iteration before the algorithm arrived at $\sigma_0'$. Therefore, to find the first vertex whose balance becomes 0 as $w$ is increased we need only to find the vertex whose balance was the smallest, among those vertices with indices higher than $j$, at the start of major iteration $i$. From Proposition 21 this is clearly the vertex with index $j+1$.

We have now fully proved the first and third parts of the proposition. To finish the proof we need only to compute the value that $w$ must be set to in order to make the vertices with index $j+1$ have a balance equal to 0. Since the balance of these vertices has always decreased, this must be equal to the balance that these vertices had at the start of major iteration $i$. By Proposition 21 this is equal to:

$$2 \times (2^i - \sum_{k=j+2}^{i-1} 2^k)$$

□

Finally, we can provide a proof for Theorem 19, by showing that the Cottle-Dantzig algorithm will take an exponential number of steps on this family of games.

*Proof (of Thoerem 19).* The algorithm first spends $n$ steps adding the Min vertices to the non-negative set. Then, in major iteration $i$, Proposition 22 leads to the same recursion that appears in the proof of Theorem 18. Thus, while increasing $w$ the algorithm must pass through $T(i-1) = 2^i - 2$ strategies. The algorithm will then switch the Max vertex with index $i$, which means that it will traverse $2^i - 1$ strategies in total during major iteration $i$. This gives the total number of iterations used by the algorithm as

$$n + \sum_{i=1}^{n}(2^i - 1) = 2^{n+1} - 1.$$

□